\documentclass[draft,usenatbib]{mn2e}
\usepackage{psfig}

\def\gsim{\mathrel{\raise0.35ex\hbox{$\scriptstyle >$}\kern-0.6em % Greater/squiggles
\lower0.40ex\hbox{{$\scriptstyle \sim$}}}}
\def\lsim{\mathrel{\raise0.35ex\hbox{$\scriptstyle <$}\kern-0.6em % Less than/squiggles
\lower0.40ex\hbox{{$\scriptstyle \sim$}}}}

\def\arcsec{\hbox{$^{\prime\prime}$}}

\def\fp{\hbox{f}_{\rm p}}
\def\fred{\hbox{f}_{\rm red}}
\def\fblue{\hbox{f}_{\rm blue}}
\def\Rc{$R_c$}
\def\Bj{$b_{\rm J}$}
\def\MBj{$M_{b_{\rm J}}$}
\def\Mstar{\hbox{M}_{*}}
\def\Halpha{$\rm H\alpha$}
\def\ewoiim{\raise0.35ex\hbox{$\scriptstyle <$}\hbox{\rm EW[OII]}\raise0.35ex\hbox{$\scriptstyle >$}}
\def\ewoiimsf{\raise0.35ex\hbox{$\scriptstyle <$}\hbox{\rm EW[OII]}{\LARGE\hbox{$\scriptstyle |$}}\hbox{SF}\raise0.35ex\hbox{$\scriptstyle >$}}
\def\kmsMpc{$\rm kms^{-1}Mpc^{-1}$}

\def\zform{$z_{\rm form}$}
\def\ztrunc{\hbox{z}_{\rm trunc}}
\def\ptrunc{\hbox{P}_{\rm trunc}}
\def\pfldgrp{\hbox{P}_{\rm f:g}}
\def\Pergr{\hbox{$\xi$}_{\rm gr}}

\def\svi {\sigma(v)_{intr}}
\def\dr {\delta(r)}

\begin{document}

\title{Galaxy Groups at ${\bf 0.3 \leq z \leq 0.55}$. II. Evolution to
  ${\bf z \sim 0}$}

\author[D.~J.~Wilman et al.]
{D.~J.~Wilman$^{1,2,7}$,~M.~L.~Balogh$^{1,3}$,~R.~G.~Bower$^1$,~J.~S.~Mulchaey$^4$,
  \and
  ~A.~Oemler~Jr$^4$,~R.~G.~Carlberg$^5$,~V.~R.~Eke$^1$,~I.~Lewis$^6$,~S.~L.~Morris$^1$,
  \and
  ~R.~J.~Whitaker$^1$\\
  $^1$Physics Department, University of Durham, South Road, Durham DH1 3LE, U.K.\\
%  $^2$Astrophysics \& Gravitation Group, 
  $^2$Max-Planck-Institut f\"ur extraterrestrische Physik, Giessenbachstra\ss e, D-85748 Garching, Germany (present address)\\
  $^3$Department of Physics,
  University of Waterloo, Waterloo, Ontario, Canada N2L 3G1 (present address).\\
  $^4$Observatories of the Carnegie Institution, 813 Santa Barbara Street, Pasadena, California, U.S.A.\\
  $^5$Department of Astronomy, University of Toronto, Toronto, ON, M5S 3H8 Canada.\\
  $^6$Department of Physics, University of Oxford, Keble Road, Oxford, UK\\
  $^7$\emph{email: dwilman@mpe.mpg.de}}

\maketitle

\begin{abstract}

We compare deep Magellan spectroscopy of 
  26 groups at $0.3 \leq z \leq 0.55$, selected from the Canadian
  Network for Observational Cosmology 2 field survey, with a large
  sample of nearby groups 
  from the 2PIGG catalogue \citep{Eke04}. We find that the fraction of
  group galaxies with significant [O{\sc ii}]$\lambda3727$ emission
  ($\geq$5\AA) increases strongly with redshift, from $\sim 29$\% in 2dFGRS to $\sim 58$\% in CNOC2, for all
  galaxies brighter than $\sim \Mstar+1.75$. 
  This trend is parallel
  to the evolution of field galaxies, where the equivalent fraction of emission
  line galaxies increases from $\sim 53$\% to $\sim
  75$\%. 
  The fraction of emission-line galaxies in groups is 
  lower than in the field,
  across the full redshift range, indicating that
  the history of star formation in groups is influenced by their
  environment. We show that the evolution required
  to explain the data is inconsistent with a quiescent model of galaxy
  evolution; instead, discrete events in which galaxies
  cease forming stars (truncation events) are required. We constrain the probability of truncation ($\ptrunc$) and
  find that a high value is required in a simple
  evolutionary scenario neglecting galaxy mergers ($\ptrunc \gsim
  0.3~$Gyr$^{-1}$). However, without assuming significant density
  evolution, $\ptrunc$ is not required to be larger in groups than in
  the field, suggesting that the environmental dependence of star formation was
  embedded at redshifts $z \gsim 0.45$.
\end{abstract}


\section{Introduction}
Star formation rates derived from high redshift UV surveys and low
redshift spectral analysis indicate that the global star formation
rate has declined since $z \sim 1.5$ by an uncertain factor of between
4.0 and 40.0
\citep[e.g.][]{Lilly96,Madau98,Wilson02,Panter03}.  This evolution is
apparently associated with downsizing
\citep{Cowie99,Kauffmann03,Poggianti03}, such that the characteristic
mass of star forming galaxies decreases with time.  The precise cause
of this decline, however, is unknown. It may be driven by internal
(local) processes, leading to the exhaustion of the gas reservoir, or
by interactions with the local environment. 
This is often referred to as the
nature versus nurture dichotomy of galaxy evolution.

The properties of a galaxy population are known to be strongly correlated
with their local environment \citep[e.g.][]{Dres80disc}: galaxies in
dense environments typically have bulge--dominated morphologies, low
star formation rates and HI gas content, and red colours. Large surveys
such as the Sloan Digital Sky Survey (SDSS) have shown that both the galaxy colour distribution
\citep[e.g.][]{Blanton03,BaloghBaldry04} and
star formation rate distribution
\citep{Lewis02,Martinez02,Gomez03,Balogh03,Kauffmann04} depend on local
galaxy density over a wide dynamic range. Similar trends have also been determined
at redshifts up to $z \sim 0.5$
\citep{Dressler97,BaloghMorr99,Kodama01,Treu03}.  
Recently, \citet{dePropris04} and \citet{BaloghBaldry04} have
shown that, while the fraction of red galaxies in the nearby Universe increases with local
density, the colour distribution and median EW[\Halpha] of the blue, star-forming galaxies
is nearly independent of environment.  A similar trend was observed in
the \Halpha\ equivalent width (EW[\Halpha]) distribution  \citep{Balogh03}.  A possible
interpretation of these trends is that dense environments transform
galaxies from blue to red on a relatively short timescale, $\lsim0.5$ Gyr.
 
In a $\Lambda$CDM Universe, the growth of large scale structure is a
consequence of the hierarchical clustering process. It is therefore possible
that this clustering process itself could drive the evolution of
global star formation, as more galaxies are drawn into dense
environments where their star formation rates are somehow suppressed.
The fraction of galaxies
located in galaxy clusters is only $\sim 10$\% even at the present
epoch and, thus, such environments alone can not have a large
influence on the global star formation rate.
However, perhaps over $50\%$ of galaxies are today found in 
groups of various sizes \citep{Eke04} and thus these environments  
may play a more significant role.  Although some proposed mechanisms
for transforming galaxies in dense environments, such as ram pressure
stripping 
\citep{GunnGott72,Quilis00} are unlikely to be effective in small
groups, many other effects, such as
strangulation
\citep{LTC,BaloghNav00,Cole00,Diaferio01},
tidal interactions \citep{Byrd90,Gnedin03}, or galaxy mergers and
interactions \citep[e.g][]{Joseph85,Moore96} may be more widespread.  In
particular, galaxy interactions 
are likely to be most common in groups, where the velocity dispersion
of the groups is not much larger than that of the constituent galaxies
\citep{Barnes85,Zabludoff98,Hashimoto00}.

Studies of nearby groups \citep[e.g.][]{Zabludoff98} show that
their galaxy
populations vary from cluster-like (mostly early types)
to field-like (mostly late-types), suggesting that a nurturing process
of galaxy evolution may well be taking place
\citep[e.g.][]{Zabludoff2000,Hashimoto00,Tran01}.   However, 
galaxy groups are inevitably much more difficult to detect than
clusters, with a relative paucity of members and significantly lower
density hot plasma.  Therefore, in
most cases the group selection criteria is either not well understood,
or biased in some way.  In particular, one
successful method has been to search for the most overdense,
compact groups \citep{Hickson82,SS01,Coziol04}; however, such systems may
be atypical of the average group environment.
Today new opportunities are afforded by large, complete catalogues of nearby groups 
compiled from redshift surveys such as SDSS and the 2dF Galaxy Redshift
Survey 
\citep[e.g.][]{Eke04}. 

One way to directly observe the influence of galaxy groups is to trace
their redshift evolution.
In rich clusters, a strong
evolution in the fraction of blue galaxies, $\fblue$, was detected by
\citet{BO84} and later by others \citep[e.g.][]{Margoniner01,DePropris03b}, although even this result is
still a matter of some debate \citep[e.g.][]{AE99,A_clusev}.
In \citet{AllSmith93}, a sample of
groups were photometrically selected in the vicinity of bright radio
galaxies at low ($z \le 0.25$) and intermediate ($0.25 \le z \le
0.50$) redshift. They tentatively confirm an analogous evolution in the fraction
of $\fblue$ in larger groups. However, the statistical limitations
of photometric data are significant, particularly through field
contamination. In addition, the radio selection might
bias the choice of groups. It is therefore important to repeat this study using
a redshift-space selected sample of spectroscopically confirmed
groups. Higher redshift catalogues of groups now exist
\citep{Cohen2000,Carlberg01} from which galaxy properties have
been analysed. For example, \citet{Carlberg01disc} studied the
properties of group galaxies in the CNOC2 group sample at intermediate
redshift. Amongst other things, they discovered a trend in
the mean galaxy colors, which on average become redder than the
field toward the group centers. In  \citet[][hereafter
Paper~I]{Wilman04a}, we present our deeper and more
complete spectroscopy in the region of the intermediate redshift CNOC2
groups \citep{Carlberg01}. Our data show that the properties of
galaxies in intermediate redshift groups are significantly different
from those of coeval field galaxies, in that group galaxies are significantly less likely
to have ongoing star formation than their field counterparts, and groups also contain a
significant excess of bright galaxies (\MBj$\leq -21$). These results
are discussed in detail in that paper.
 
In this paper, we contrast the properties of our intermediate redshift
group sample \citep[][PaperI]{Carlberg01} with a large sample of
galaxy groups at low redshift, selected from the 2dF Galaxy Redshift
Survey \citep[2dFGRS,][]{Eke04}. This
allows us to examine the evolution of galaxies in the group
environment with purely spectroscopic data and over a significant
range of redshift.  In Section~\ref{sec-data} we introduce our galaxy
and group samples at intermediate redshift (CNOC2 - see also PaperI)
and locally (2dFGRS). We then go on to ensure a fair comparison
between these two populations and the surrounding field by examining
the luminosity functions and EW[OII] distributions.  In
Section~\ref{sec-results}, we present our results, in which we assess
the environmental and evolutionary dependencies of EW[OII] as well as
the dependence on other parameters such as galaxy luminosity.  We then
discuss the scientific implications in
Section~\ref{sec-discussion}, and present simple models for the 
star formation history of these galaxy populations to constrain theories
of galaxy evolution. Section~\ref{sec-conclusions} presents our final
conclusions.

Throughout this paper we assume a $\Lambda$CDM cosmology of
$\Omega_{M} = 0.3$, $\Omega_{\Lambda}=0.7$ and $H_0 = 75$\kmsMpc.

\section{Data}\label{sec-data}

\subsection{CNOC2: The Intermediate Redshift Sample}

A complete description of the data and reduction methods can be found
in Paper~I.
In summary, the intermediate redshift group sample is selected from
the CNOC2 redshift survey in the range $0.1\leq z \leq 0.55$
\citep{Carlberg01,Yee00}. 
We obtained deeper, more complete
multi-object-spectroscopy in the regions of 26 of these groups (in 20
fields) at
$0.3\leq z \leq 0.55$ using LDSS2 on the 6.5m Baade telescope at Las
Companas Observatory in Chile. The fields were chosen to maximize the
number of groups in the \citet{Carlberg01} sample along the line of
sight, within this redshift range. 
Redshifts were measured for $74\%$ of galaxies targetted with
\Rc$\leq 22$ (with $\sim 60\%$ success rate in the faintest bin $21.5
< $\Rc$ \leq 22$) and galaxies have been reassigned to groups with a
new determination of the group velocity dispersion. The galaxies have
each been weighted by a factor $W_{C}$ to account for radial and magnitude-dependent
selection functions (see Paper~I).

The CNOC2 field galaxy sample is defined to include all galaxies
within $240\arcsec$ 
of a targetted group centre,
lying within the redshift range
$0.3 \leq z \leq 0.55$ but excluding those galaxies assigned to the targetted
group to 
avoid biasing the field towards the group environment. The
final magnitude--limited field sample is therefore representative of the Universe in the
$0.3 \leq z \leq 0.55$ redshift range.

The CNOC2 group sample contains 240 galaxies within $1h_{75}^{-1}$Mpc
of the group centre and the field sample contains 334 galaxies.

\subsection{2dFGRS: The Local Sample}

The local redshift galaxy sample comes from the large 2dFGRS  with over
220 000 galaxy spectra 
selected in the \Bj-band. The galaxy sample is effectively volume
limited (with low incompleteness the sample is representative of the
whole population) in the redshift range $0.05 \leq z \leq 0.1$ for
galaxies with \MBj$ \leq -18.85$.  Although there were problems with
the atmospheric dispersion corrector prior to August 1999 which affect
the instrument throughput \citep{Lewis02_inst}, we find our
results are unchanged if we exclude data obtained in this period.  

The 2dFGRS Percolation-Inferred Galaxy Group catalogue
\citep[2PIGG][]{Eke04} is also based on a friends-of-friends
percolation algorithm. An axial ratio
(defined as the line-of-sight~length relative to the projected~spatial~length) of $\sim11$
is used to link 2dFGRS galaxies together,
forming a large catalogue of local groups. Velocity
dispersions of the 2PIGG (and CNOC2) groups are calculated with the
gapper algorithm. Full details of the 2PIGG group-finding algorithm and
description of the catalogue can be found in \citet{Eke04}. We only
investigate groups with number of known members $N_{m} \geq 10$
because the contamination of that group catalogue with unphysical
systems becomes large in smaller groups. We note that the CNOC2
group detection
algorithm \citep{Carlberg01} requires more bright members in close proximity to
each other, and therefore likely suffers from
less contamination. Even with the $N_{m} \geq 10$ requirement for
2dFGRS groups, we find that the range of group velocity dispersion
matches that seen in the CNOC2 group sample. From now on we will refer
to the 2dFGRS sample simply as the \emph{2dF sample}.

The 2dF field is defined as all galaxies in the 2dF galaxy
catalogue and represents the global galaxy population in the $0.05\leq
z \leq 0.1$ redshift range, within magnitude limits. Since these
groups were untargetted, this definition is compatible with our CNOC2
field definition.

The 2dF group sample contains 5490 galaxies within $1h_{75}^{-1}$Mpc
of the group centre and the field sample contains 50981 galaxies.

\subsection{The galaxy luminosity function in groups and the field}

\subsubsection{K-corrected rest frame magnitudes}
Galaxies in both catalogues are k-corrected to give rest-frame
absolute magnitudes in the \Bj-band. The 2dF k-corrections are taken 
from \citet{Norberg02}, and are generally small.  For the CNOC2 survey,
k-corrections have
been calculated using no-evolution models; 
we will explore the
sensitivity of our results to the assumed model of galaxy evolution in
the discussion (Section~\ref{sec-discussion}).  We note
that k+e corrections are also available for galaxies in the original
CNOC2 sample \citep{Shepherd01}.  For each
galaxy, we first choose a mixture of observed local SEDs \citep{KE85}
for which the model B-I colour matches the observed colour at the
given redshift. Then the rest-frame absolute \Bj\ magnitude can be
determined from the models, given the observed \Rc\ magnitude, SED
mixture and redshift. The transformation from observed \Rc-band
magnitude to rest-frame \Bj\ is chosen because it is closely matched
at CNOC2 redshifts and because it directly transforms the CNOC2
spectroscopic selection band to the 2dF selection band. Luminosities
are then corrected for galactic extinction on a patch-to-patch basis,
computed by extrapolating from B and V band extinction values obtained
from NED \citep[][variation within each patch is
negligible]{Schlegel98}. We make no correction for internal extinction, also to
allow direct comparison with local galaxies in 2dF.

\begin{figure}
  \centerline{\psfig{figure=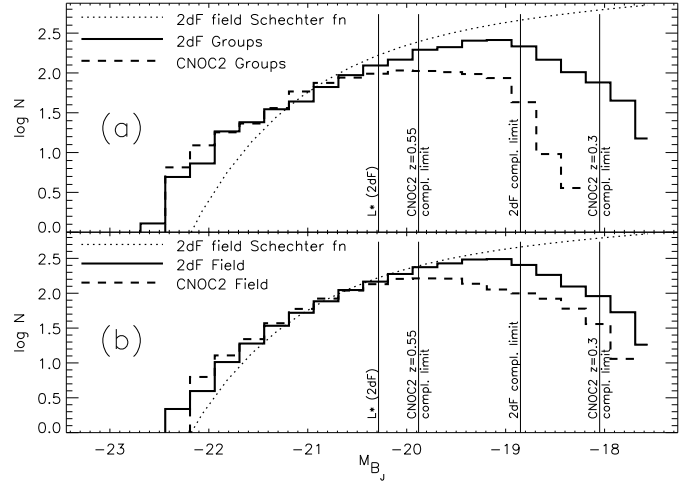,width=0.5\textwidth}}
  \caption{Luminosity functions of the group galaxies ({\bf a}) and field
    galaxies ({\bf b}) for 2dF and CNOC2. Field and Group CNOC2 data are
    each normalised to match the total number in the CNOC2 sample for
    \MBj$<-20.25$} 
  \label{fig:lumfns}
\end{figure}

\subsubsection{Luminosity limits}
Figure~\ref{fig:lumfns} shows the superimposed luminosity functions of
the 2dF and CNOC2 group and field samples. The volume--limited 2dF
sample is $>90\%$ complete for \MBj$\leq -19$ and so we apply no
completeness correction. The Schechter function computed for the 2dF survey
\citep{Norberg02} 
is shown for comparison.
The CNOC2
galaxies are weighted by $W_{C}$ to correct for the selection
function. For comparison between 2dF and CNOC2 group and field
samples, the data are normalised so that there is the same number of
weighted galaxies brighter than \MBj$=-20.25$. At these magnitudes,
neither sample suffers any incompleteness due to falling below the
apparent magnitude limit in the redshift range considered.
We note that the enhanced bright to faint
galaxy ratio seen in CNOC2 groups relative to the field (Paper~I) is
also seen in the local 2dF groups.

Also shown in Figure~\ref{fig:lumfns} are some critical values of
luminosity. The value
of $\Mstar$ in \citet{Norberg02}, appropriate to our cosmological model, is $\sim 
-20.28$, and the 2dF data are complete down to $-18.85$, or equivalently
$\sim \Mstar+1.5$. Our CNOC2 data 
span a wide range in redshift and thus the luminosity limit
corresponding to our apparent magnitude limit of \Rc=22 is 
redshift dependent. At the upper limit of our redshift range,
$z=0.55$, a \Rc$=22$ galaxy with a mean k-correction will transform to
a rest-frame luminosity \MBj$=-19.75$ and at the lower redshift limit
of $z=0.3$, the same galaxy would transform to \MBj$= -17.93$. In the
case of the reddest galaxies with larger K-corrections, these limits
would lie at \MBj$=-20.07$ ($z=0.55$) and \MBj$=-18.06$ ($z=0.3$), so
we are incomplete below these magnitudes. 

Most galaxies in our CNOC2 and 2dF catalogues lie below the
brightest CNOC2 luminosity limit of \MBj$=-20.07$. To enable us to 
compare the 2dF and CNOC2 galaxy samples independently of differences
in the luminosity function (which may be partly intrinsic but is mostly
due to selection effects), we choose to
apply an additional luminosity weighting to the CNOC2 galaxies.
This weighting is calculated within each bin in luminosity using the
formula: \begin{equation} W_{lum} = N_{2dF}/\sum_{i =
    1}^{N_{CNOC2}}W_{C} \end{equation} which corresponds to the
difference between the field luminosity functions. It is
applied in the range $-20.25 \leq $\MBj$ \leq -18.5$, where the CNOC2
data become incomplete at the high redshift end. 
The choice of a faint final
luminosity limit of \MBj$ = -18.5$ ($\sim \Mstar+1.75$)
makes maximal use of the data and allows the properties of faint
galaxies to be compared with those of brighter galaxies. We emphasize
that whilst we are incomplete at \MBj$\gsim -19.75$ at $z=0.55$ in
CNOC2 and \MBj$\gsim -18.85$ in 2dFGRS, this has no impact on any analysis
of galaxy properties as a function of luminosity or on comparisons
between the group and field galaxy populations. Also, when studying
galaxy properties as a function of luminosity, the analysis is
independent of the CNOC2 galaxy weighting, including little effect from
weighting by the selection function.

\subsection{Measurement of star formation using EW[OII]}
The \Halpha\ 
emission line disappears entirely from the LDSS2 spectrograph window
at $z>0.21$, limited by the  instrument sensitivity of the
current optics and detector. Therefore,we use the [OII]$\lambda3727$ emission line
equivalent width (EW[OII])
to study the relative levels of star formation in our galaxy
samples.  In Paper~I we outlined the reasons why EW[OII] is sufficient to reveal trends of star formation with
galaxy environment.  
In contrast to using the line flux, the effect of
normalising by the continuum when computing the equivalent width reduces uncertainties
related to absorption by dust and aperture bias, which are relevant when
comparing galaxy properties at different redshifts. In particular, we
show in Section~\ref{sec-apbias} that our analysis is
insensitive to aperture bias in EW[OII].  

\subsubsection{Fair comparison of EW[OII] in 2dF and CNOC2}
\begin{figure}
  \centerline{\psfig{figure=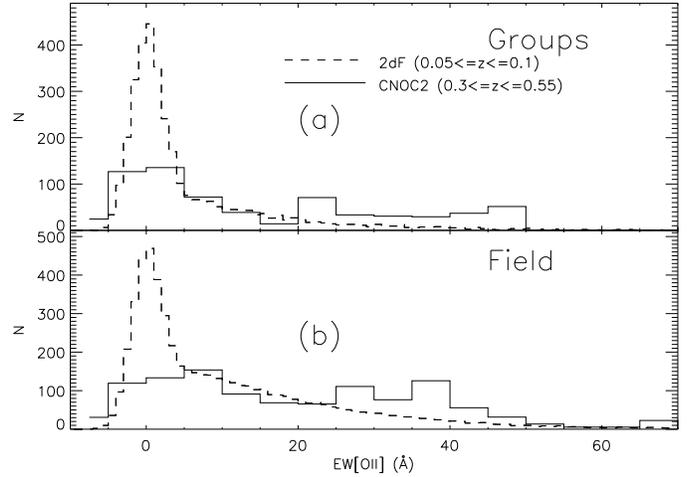,width=0.5\textwidth}}
  \caption{Normalised distributions of EW[OII] of the group ({\bf a}) and field
    ({\bf b}) galaxies in 2dF ($0.05\leq z \leq 0.1$) and CNOC2 ($0.3 \leq z
    \leq 0.55$) samples where \MBj$\leq -18.5$. The data is normalised
    to match the total number of galaxies in the CNOC2 analaysis.} 
  \label{fig:oiidistribs}
\end{figure}

Details of the CNOC2 equivalent width measurement process are given in
Paper~I. In 2dFGRS, the equivalent widths of [OII] are measured in a
similar way to \Halpha\ using a completely automated fitting
procedure, \citep[see][for details]{Lewis02}. In the fitting of the
[OII] emission line, many 2dF measurements are classified as \emph{no
  line present}. In our analysis, these are set to 0\AA\ and then
all 2dF measurements are smoothed with a gaussian kernel of
width 2\AA\ to match the mean error on CNOC2 EW[OII] measurements
(much greater than the 2dF line measurement error of $\ll 1$\AA). We
note that the fraction of galaxies with EW[OII]$>$5\AA\ is unchanged
by this smoothing to within $<1\%$. Figure~\ref{fig:oiidistribs} shows the
distribution of EW[OII] in our 2dF and CNOC2 group and field galaxy
catalogues.
We limit the group data to within 
$1 h_{75}^{-1}$Mpc (projected) of the group centre in all cases. The
CNOC2 galaxies 
are weighted by a combined completeness and luminosity weighting to
match the 2dF luminosity function, $W_{tot} = W_{C}\times W_{lum}$.
The CNOC2 galaxies are limited to \Rc$\leq 22$ and all galaxies are
limited to \MBj$\leq -18.5$. Finally, the distribution of 2dF EW[OII]
is normalised to provide an equal number of galaxies to that found in
CNOC2, for presentation only. This is done independently for the group
and field populations.

\subsubsection{Diagnostics of Star Formation for a Galaxy Population}

We are motivated by the findings of
\citet{Strateva01,Blanton03,Baldry03} and \citet{Balogh03} who show
that galaxy populations have a bimodal
distribution in colour and EW[\Halpha].
\citet{Balogh03,BaloghBaldry04} show that the fraction of red,
  passive galaxies is strongly dependent upon local galaxy density.
The division between passive and star forming galaxies in the
EW[\Halpha] distribution occurs at $\sim 4$\AA\ \citep{Balogh03}. We
do not expect to see such a clear bimodality in EW[OII] since 
EW[\Halpha] $= 4$\AA\ typically corresponds to EW[OII] $< 2$\AA, below the
measurement error in EW[OII] for CNOC2. Greater intrinsic scatter in
the SFR-[OII] relation than in the SFR-\Halpha\ relation also works to
mask the division between the two populations. 
Thus, although we cannot cleanly separate the two populations, we impose
an arbitrary division at 5\AA\
in the CNOC2 and smoothed 2dF data.
We expect the
population with EW[OII]$<$5\AA\ to be dominated by the passive
population\footnote{We note that the
shape of the negative
side of the 0\AA\ peak in the EW[OII] distribution from the full CNOC2
survey is consistent with a gaussian function, supporting the
hypothesis that this peak is dominated by 
galaxies with no [OII] emission and normally distributed errors
\citep{Whitaker04}.}  and the 
population with EW[OII]$\geq$5\AA\ to be dominated by the star-forming
population, and this division is sufficient to reveal trends in
the data \citep[see e.g.][]{CFRSXIV97,Zabludoff2000}. 

To assess the relative normalisation of the two
populations, we define {\bf$\fp$} as the fraction of passive
    galaxies. The level of [OII] emission in the star forming
galaxies is characterised by {\bf $\ewoiimsf$}, which represents the 
  median EW[OII] restricted to star--forming galaxies.

We note that it will also be interesting to derive the fraction of blue galaxies using CNOC2 colours, for a more direct comparison with classical studies of the Butcher-Oemler effect. However, this analysis is not straightforward, because of the complex dependence of CNOC2 colour apertures on galaxy size, galaxy type and redshift; the meagerness of the group red sequence and the difficulties in making a direct comparison with 2dFGRS. Many of these problems can be overcome by computing the fraction of galaxies in each peak of a bimodal colour distribution. This analysis will be presented in a future paper, currently in preparation.

\subsubsection{Aperture bias}\label{sec-apbias}
Systematic effects on the measurements of EW[OII] can be induced by
the relative aperture sizes used in
the 2dFGRS and CNOC2 spectroscopy.  In particular, the 2dF fibres
generally sample light from a smaller physical radius than the CNOC2
slits, and this might lead to an overestimate of $\fp$ by excluding the
star forming 
regions in face-on disk galaxies. In Appendix~\ref{sec-apeffects}, we
use SDSS resolved photometry to estimate the effects of aperture bias
across our magnitude range. 
We find that the fraction of
galaxies found in the red peak of the bimodal colour
distribution is no greater when considering colours measured
inside $3\arcsec$ SDSS fibres, rather than the total colour.
This is because both red and
blue galaxies have similar colour gradients, likely due to metallicity
rather than star formation. 
Thus we conclude that the effects of aperture bias do not strongly
affect our measurements of $\fp$.

\section{Results}\label{sec-results}

\subsection{Evolutionary and Environmental Dependencies of EW[OII]}

A comparison of the 2dF and CNOC2 EW[OII] distributions in
Figure~\ref{fig:oiidistribs} shows that the fraction of galaxies in
the 0\AA\ peak depends on both epoch and environment.  In particular,
the 0\AA\ peak in the 2dF data is much more prominent than in
the CNOC2 survey, for both field and groups.  At both epochs, however,
the group galaxy population is more
biased towards the 0\AA\ peak than the corresponding field population.
We now explore these trends in more detail.

\subsubsection{The dependence of $\fp$ on redshift, environment and luminosity}

\begin{figure}
  \centerline{\psfig{figure=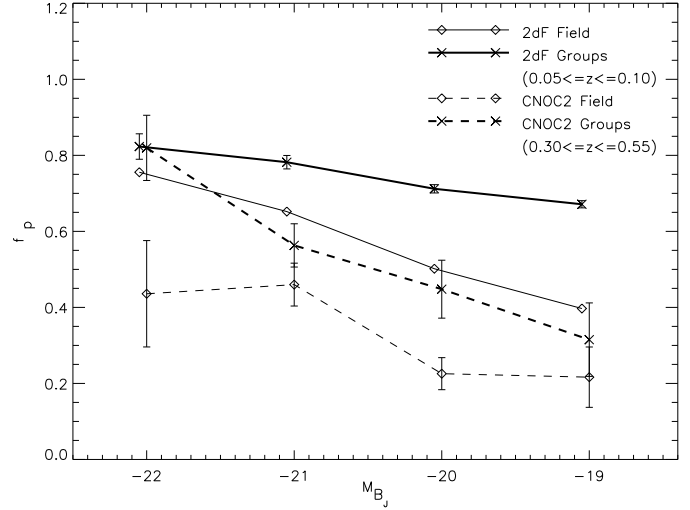,width=0.5\textwidth}}
  \caption{The fraction of galaxies with EW[OII]$<$5\AA\ (the passive
    population), $\fp$, in 2dF and CNOC2 groups (within 1 projected
    Mpc of group centre) and field, as a function of \MBj. The 2dF
    points are slightly offset in \MBj for clarity.} 
  \label{fig:fracSFLum}
\end{figure}

In Figure~\ref{fig:fracSFLum}, the fraction of passive galaxies, $\fp$
is plotted against rest-frame \MBj luminosity. Statistical errors are
estimated using a Jackknife resampling method \citep{Jackknife}. 
We
can see that:

\begin{itemize}
  
\item{In all samples, $\fp$ is a strong function of luminosity with
    fainter galaxies far more likely to be star forming than brighter
    galaxies at equivalent redshifts. This is consistent with many
    previous results, e.g. \citet{Kauffmann03}. }
  
\item{$\fp$ is significantly greater in the galaxy groups than in the
    field at both low and intermediate redshift and also right across
    the luminosity range investigated.}
  
\item{$\fp$ is strongly redshift dependent, both in the field and in
    galaxy groups. At brighter magnitudes than \MBj$=-18.5$ 
($\sim \Mstar+1.75$ in 2dF), $\fp$ in groups evolves from $\sim 42\%$
at $0.3 \leq z \leq 0.55$ in CNOC2 to $\sim 71\%$ at $0.05 \leq z \leq
0.1$ in 2dF. In our field samples (defined to represent the global
population), $\fp$ evolves from $\sim 25$\% to $\sim 47$\% over the
same redshift interval. The observed field evolution is consistent
with the equivalent strong evolution in the fraction of passive
galaxies in the Canada-France Redshift Survey \citep{CFRSXIV97} and
the global decline in star formation rate since $z \sim 1$
\citep[e.g.][]{Madau98}. We refer to Whitaker et al., 2004, in
preparation, for a more detailed and thorough discussion of the
evolution of star formation rate in the global CNOC2
population. This provides
    a clear analogy at lower densities to the observed evolution of
    the blue galaxy fraction in clusters \citep{BO84} and to similar
    evolution in rich groups, estimated by \citet{AllSmith93} using
    photometric data.}

\end{itemize}

\subsubsection{The properties of the star forming population}

\begin{figure}
  \centerline{\psfig{figure=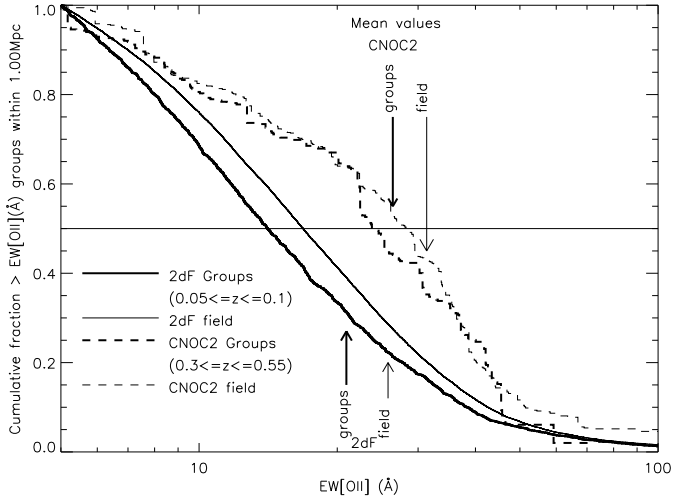,width=0.5\textwidth}}
  \caption{Cumulative  distributions of EW[OII] in star forming
    galaxies in the field and group galaxies of 2dF ($0.05\leq z \leq
    0.1$) and CNOC2 ($0.3 \leq z \leq 0.55$) samples. The arrows
    indicate the mean values of EW[OII] for star forming galaxies
    in each sample.} 
  \label{fig:CumoiiSFdistribs}
\end{figure}

\begin{figure}

  \centerline{\psfig{figure=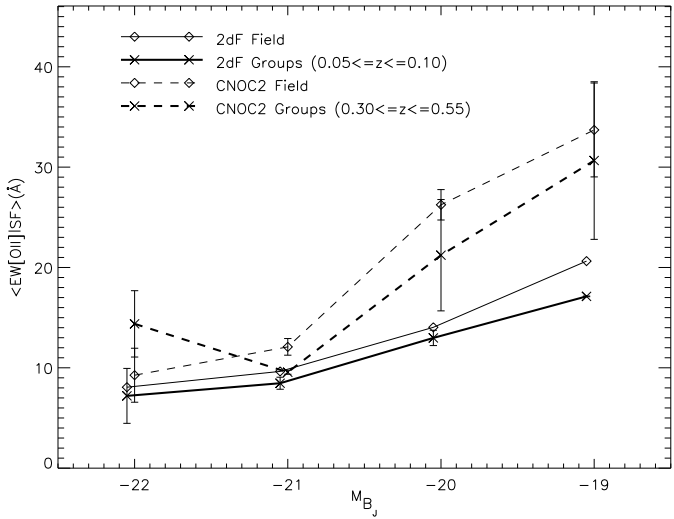,width=0.5\textwidth}}
  \caption{The median EW[OII] (\AA) of star-forming
      galaxies, $\ewoiimsf$, in the 2dF and CNOC2 galaxies in the
    groups (within 1 projected Mpc of group centre) and field, as a
    function of \MBj. The 2dF points are slightly offset in \MBj for
    clarity.} 
  \label{fig:wtmedianSFLum} 
\end{figure}

Figure ~\ref{fig:CumoiiSFdistribs} shows the cumulative distribution
of EW[OII], for star forming galaxies only, in the 2dF and CNOC2
group and field samples. Interestingly, the shape of the distribution
at a particular epoch (i.e. 2dF or CNOC2) is approximately independent of
environment, consistent with earlier analysis of the EW[\Halpha] distribution
\citep{Balogh03}. In the CNOC2 sample, the distribution shows a small
enhancement in highly star forming galaxies (EW[OII] $\gsim$
30\AA) relative to the 2dF galaxies in both groups and the field. The
mean values of EW[OII] for each sample are indicated by the arrows in
Figure~\ref{fig:CumoiiSFdistribs}. These values of 26.4\AA\ (groups)
and 31.3\AA\ (field) in CNOC2 are within $\sim 20$ per cent of the mean
values in the 2dF (20.9\AA\ in groups and 25.8\AA\ in the field) in
2dF.  This difference is much smaller than the 
evolution in $\fp$, which is almost a factor of two. 
Therefore, the evolution
of the total star formation rate is  driven more by the evolution of
$\fp$ than by evolution of the mean properties of star forming
galaxies.  

In Figure~\ref{fig:wtmedianSFLum} we show the median EW[OII] among
star--forming galaxies, $\ewoiimsf$, as a
function of luminosity in the CNOC2 and 2dF group and field samples.
Errors are again computed using a Jackknife resampling method. 
The median EW[OII] is significantly larger for fainter galaxies 
than for bright galaxies. Furthermore, the evolution in the EW[OII]
distribution is
largely manifested as an increase in $\ewoiimsf$
in galaxies with \MBj$\gsim -21.5$. 

\subsubsection{Dependence of $\fp$ in groups upon group-centric radius and velocity dispersion}

\begin{figure}
  \centerline{\psfig{figure=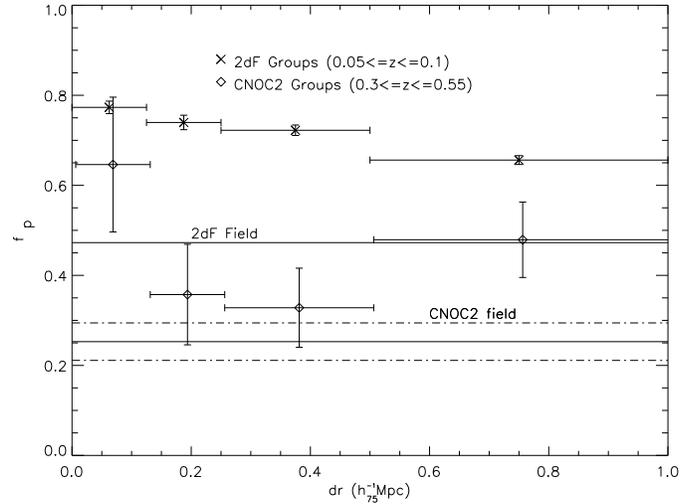,width=0.5\textwidth}}
  \caption{The fraction of galaxies with EW[OII]$<$5\AA\ (the passive
    population) ($\fp$) in 2dF and CNOC2 groups, as a function of
    physical radius. The field level is overplotted and only galaxies
    where \MBj$\leq -18.5$ are considered.} 
  \label{fig:fracSFrad}
\end{figure}

In Figure~\ref{fig:fracSFrad} we plot $\fp$ as a function of the
projected physical distance, dr, from the group centre. It is clear
that even in the better sampled 2dF groups, the total fraction of
passive galaxies, $\fp$ merely declines from $\sim0.76$ in the
innermost $0.125~h_{75}^{-1}$~Mpc bin to $\sim0.65$ in the $0.5~h_{75}^{-1}$~Mpc$ < \dr \leq
1~h_{75}^{-1}$~Mpc bin. This is a weak, but statistically significant trend. The
value of $\fp$ for the total 2dF field population is $\sim0.47$, much
lower than in the groups. Similarly, in the CNOC2 field, $\fp = 0.25$
is much smaller than that in the combined group population where $\fp
= 0.42$. We only see a trend with dr in the inner regions of the CNOC2 group population, as shown
in Paper~I. However, a trend as weak as that seen in 2dF group galaxies
would be masked by the statistical errors.

\begin{figure}
  \centerline{\psfig{figure=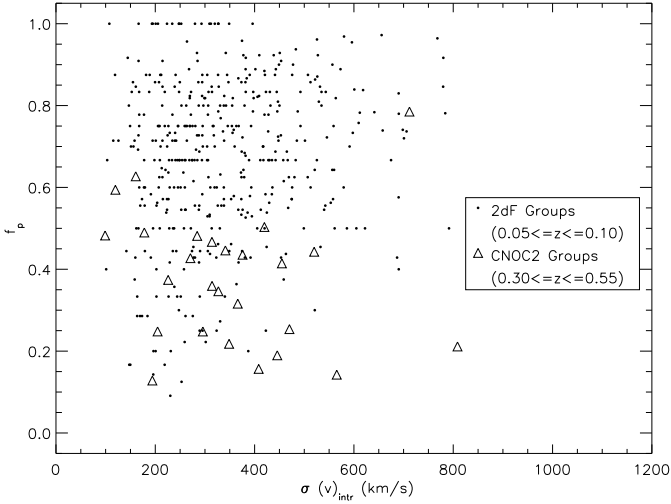,width=0.5\textwidth}}
  \caption{Fraction of galaxies with EW[OII]$<$5\AA\ (passive galaxies) ($\fp$) in 2dF and CNOC2 groups, as a function of group line of sight velocity dispersion. The points represent 2dF groups and the triangles represent CNOC2 groups.}
  \label{fig:fracSFveldisp}
\end{figure}

We have also investigated how the star formation properties of galaxy groups\footnote{We note
that whilst $\fp$ in CNOC2 groups is computed with each galaxy
weighted by its combined completeness and luminosity weighting
$W_{tot}$ (strictly only applicable when the full stacked group is
considered) we find that $\fp$ is insensitive to this weighting and
very close to the value obtained with no weighting applied.} 
depend upon the group velocity dispersion, $\svi$. In Paper~I we found
that there is little dependence of $\fp$ on $\svi$ when computed over
a wide range of galaxy luminosity. Figure~\ref{fig:fracSFveldisp}
shows $\fp$ in CNOC2 and 2dF groups as a function of $\svi$. There are
no clear trends visible in Figure~\ref{fig:fracSFveldisp} other than
the evolution from lower $\fp$ in CNOC2 groups to that seen in 2dF
groups. It is also noticeable that there are very few 2dF groups with
high velocity dispersion ($>400$km~s$^{-1}$) and low $\fp$ ($<0.4$), which is
a common regime for CNOC2 groups. We note that the enhancement of
$\fp$ in CNOC2 groups over the CNOC2 field holds when we exclude the 2
groups with velocity dispersion $>600$km~s$^{-1}$ (for more details see
Paper~I). The only CNOC2 group with $\fp > 0.7$ (group 138)
is
characterized by a high number of confirmed members (35 compared to 19
in the group with the next highest $\svi$) and might better be considered a poor cluster \citep[see
e.g.][for the behaviour of $\fp$ in clusters]{Nakata04}.

\section{Discussion}\label{sec-discussion}

\subsection{Implications}

We have detected a strong evolution in the fraction of passive
galaxies, $\fp$ (defined as the fraction of galaxies with EW[OII]$<$
5\AA), in both groups and the field since the Universe was $\sim
1/3$ its current age. Thus, the Butcher--Oemler effect
\citep[e.g.][]{BO84,Margoniner01,DePropris03b} is not strictly
a cluster phenomenon, but is seen both in the general field and in
small groups, as previously found by \citet{AllSmith93}.
These results represent clear evidence that a significant proportion
of galaxies in both environments have ceased forming stars since $z\sim
0.5$.  
A simple argument, neglecting density and luminosity
evolution, sees 50\% of star forming group galaxies and 35\% of star
forming field galaxies at $z\sim 0.4$ (CNOC2) becoming passive galaxies by
$z\sim 0.1$ (2dF). This evolution is modelled in more detail in
Section~\ref{sec-models}.

In contrast, while we find significant redshift evolution in the shape of
the EW[OII] distribution for star-forming galaxies, there is little or
no dependence on environment.
This suggests that this evolution results from very local
(i.e. internal to the galaxy) processes that drive an evolution in SFR
or metallicity, rather than external,
environmental influences.
We caution however that amongst star-forming
galaxies, it is not possible with current data to rule out an aperture
bias in the measurement of EW[OII], which might lead to
underestimation of EW[OII] in 2dFGRS star forming galaxies.

In the following sections, we make a first attempt to quantitatively decouple the environmental
dependence of galaxy evolution from the global SFR evolution. 
To fully understand this, we require a homogeneous sample over a large
range of environments.
We are currently working on providing a fair comparison
study in cluster cores \citep{Nakata04} using spectroscopic data over
an equivalent redshift range. Understanding the importance of galaxy
evolution in groups with respect to cluster cores will help to isolate
the environments in which galaxy evolution is most active. 
Complementary studies of the evolution
in isolated galaxies would be of especially great benefit to
understand the role of environment in driving galaxy evolution, and in
particular, the evolution of $\fp$.

\subsection{Modelling and Interpretation}\label{sec-models}

In this Section, we show how the strong evolution in $\fp$ seen in our
results can be interpreted in the context of galaxy evolution models.
By combining the \citet{BC03} models of luminosity evolution with a
range of star formation histories, we attempt to recreate a realistic
evolution scenario which can reproduce the observed evolution.

We assume that CNOC2 galaxies represent a population equivalent to the
progenitors of the 2dF population; therefore, using the stellar
population models of \citet{BC03} it is possible to create mock-2dF
populations by evolving the CNOC2 galaxies to the mean redshift of 2dF
galaxies ($z_{2dF}$ = 0.08) in accordance with a chosen set of model
parameters. We present two model methods with different evolutionary
scenarios but similar basic methodology. Both models allow us to
estimate the evolution of EW[OII] and \MBj within a given set of
parameters. The \emph{quiescent evolution} scenario is characterized
by the lack of environmental evolution. There are no sudden events
which drastically alter a galaxy's star formation. Bright star-forming
galaxies simply decline exponentially in their star formation with a
constant e-folding timescale and thus fade in luminosity. In the
\emph{truncation} scenario, we incorporate into our evolution model a
probability of each galaxy undergoing a truncation event, in which it
suddenly ceases star formation. In this model, there is also a
probability that a high redshift field galaxy can infall onto a galaxy
group to become a local group galaxy.

\subsubsection{The Quiescent Evolution Scenario}\label{sec-disquiesevol}

\begin{figure}

  \centerline{\psfig{figure=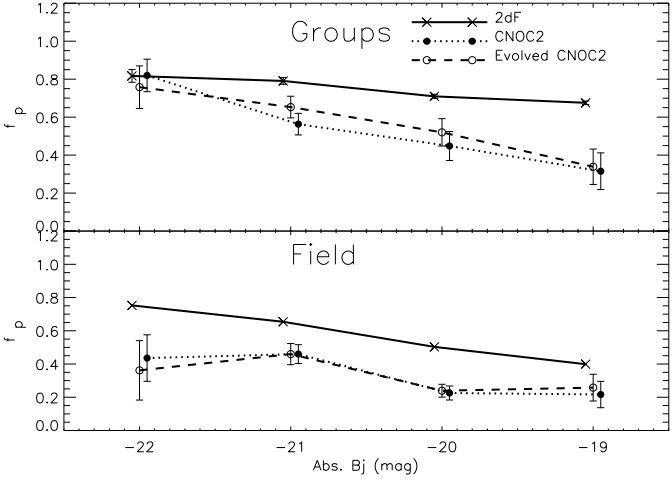,width=0.5\textwidth}}
 \caption{The fraction of passive galaxies, $\fp$, in the 2dF (solid line) and CNOC2 group (within 1
   projected $h_{75}^{-1}$Mpc of group centre) and field galaxy
   populations, as a function of \MBj. The CNOC2 galaxies are shown
   as observed (dotted-line), and evolved to $z_{2dF}$ using a quiescent
   model with a Kennicutt IMF, \zform = 10, solar metallicity and dust
   extinction (dashed-line). The 2dF points are slightly negatively
   offset, and the CNOC2 data points slightly positively offset in \MBj
   for clarity.} 
  \label{fig:PLEsensible}
\end{figure}

\begin{figure}

  \centerline{\psfig{figure=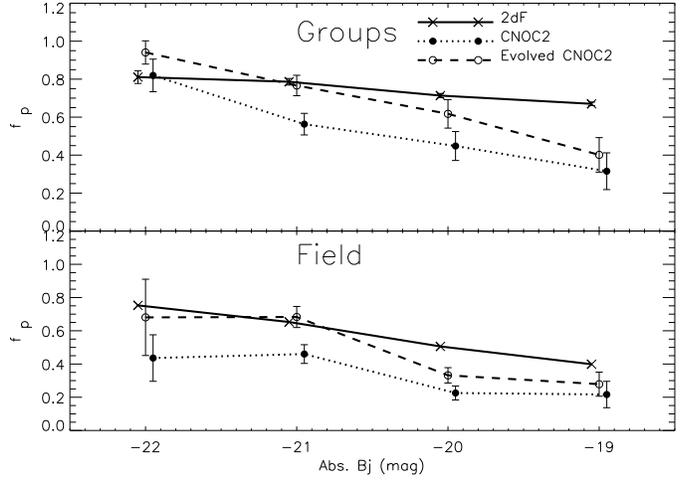,width=0.5\textwidth}}
 \caption{Similar to Figure~\ref{fig:PLEsensible},
   but assuming an extreme quiescent evolution
   model for the CNOC2 population, with a Salpeter IMF, \zform = 3, solar metallicity and no dust
   extinction (dashed-line). This represents the maximum evolution
   expected from any realistic quiescent model. The 2dF points are
   slightly negatively offset, and the CNOC2 data points slightly
   positively offset in \MBj for clarity.} 
  \label{fig:PLEextreme}
\end{figure}

Quiescent evolution describes the evolution of galaxies in which every
galaxy's star formation declines over its lifetime with a single
e-folding timescale. The modelling procedure for quiescent evolution
is described in Appendix~\ref{sec-quiesevol}, which also describes the
effects of allowing each parameter to vary.  To investigate the
effects of quiescent evolution on $\fp$, we choose a control model and
an extreme model. For our control model we choose a Kennicutt IMF, a
galaxy formation redshift \zform = 10 and solar metallicity. We also
incorporate a two-component dust prescription into the model
\citep{Granato00}. Figure~\ref{fig:PLEsensible} shows $\fp$ as a
function of luminosity in the observed CNOC2 and 2dF samples, together  with
the equivalent trend in the evolved CNOC2 population obtained
using this model. The evolution of $\fp$ in the model
significantly underestimates the trend seen in the real data. This is
partly because galaxies
which become passive also tend to fade into a fainter bin of
luminosity, leaving the trend of $f_p$ with luminosity approximately
unchanged. 
For our extreme model, we deliberately choose 
parameters which maximize the evolution in $\fp$, as discussed in
Appendix~\ref{sec-quiesevol}. We choose a Salpeter IMF, \zform = 3,
solar metallicity and we ignore the effects of dust.
Figure~\ref{fig:PLEextreme} shows the same 2dF and CNOC2 data as in
Figure~\ref{fig:PLEsensible} but this time overplotted with the
evolved CNOC2 population obtained using this extreme model. In this case,
galaxies fainter than $\sim \Mstar-0.5$ still show a
significant deficit of passive galaxies (low $\fp$) in the evolved
CNOC2 galaxies when compared to the data. This provides strong
  evidence that transformations are required to reproduce the observed
  evolution in $\fp$ both in groups and the global (field) population.

\subsubsection{The Truncation Scenario}
Next we consider a model in which
galaxies undergo transformations that cause them to 
cease forming stars. We have shown in
Section~\ref{sec-disquiesevol} that some form of galaxy
transformation appears to be required to reproduce our observed
evolution in $\fp$. Here, we constrain the probability of these
transformations required to match the observed evolution in $\fp$ and
its dependence upon galaxy luminosity and environment. The modelling
procedure is described in Appendix~\ref{sec-truncevol}.  In this
scenario, a CNOC2 galaxy either continues with its e-folding decline
in star formation, or has its star formation truncated instantaneously
with a probability $\ptrunc$ per Gyr, at a random point during its
evolution to $z_{2dF} = 0.08$. The timescale of the transformation is
likely to have little effect on the evolution of $\fp$. We neglect the
possibility that some transformations are accompanied with a strong
starburst or involve the merging of galaxies as either of these
possibilities cannot be constrained by simply considering the
evolution in $\fp$. In a later paper we will consider our data in the
context of a more complete galaxy formation model \citep[e.g.][]{Cole00}.

\begin{figure}

  \centerline{\psfig{figure=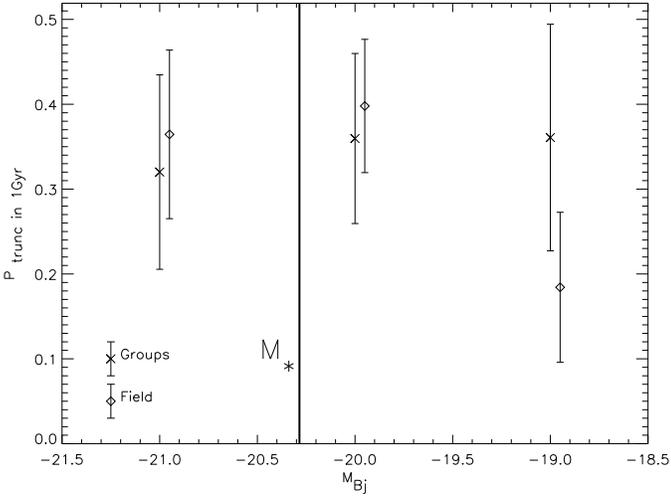,width=0.5\textwidth}}
 \caption{The probability (per Gyr) that star formation has been
   truncated ($\ptrunc$) as a function of \MBj\, as determined by
   modelling the evolution of $\fp$ between the CNOC2 ($0.3 \leq
   z_{CNOC2} \leq 0.55$) and 2dF ($0.05 \leq z_{2dF} \leq 0.1$)
   samples. The crosses represent the group galaxies and the diamonds
   represent the field population which has been artificially offset by
   0.05mag for clarity. Here, there is assumed to be no density
   evolution, i.e. all group galaxies were already in groups by
   $z_{CNOC2}$ ($\Pergr = 100$\%). The vertical line represents the
   location of a $\Mstar$ galaxy in 2dF as determined from the
   luminosity function of \citet{Norberg02}.} 
  \label{fig:ptrunclumnodensevol} 
\end{figure}

\begin{figure}

  \centerline{\psfig{figure=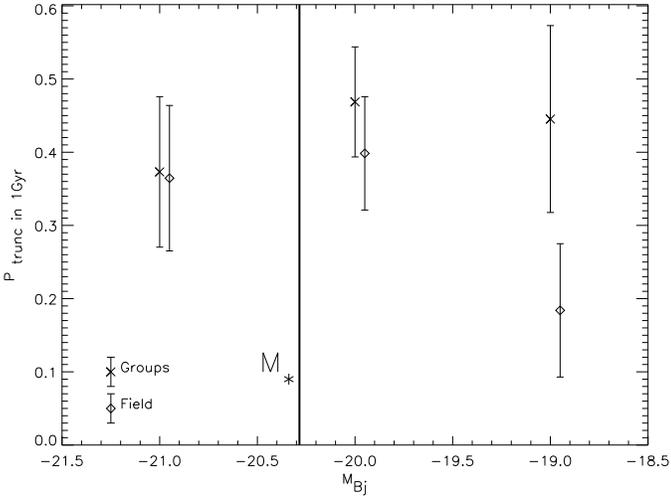,width=0.5\textwidth}}
 \caption{As Figure ~\ref{fig:ptrunclumnodensevol}, but assuming
   local 2dF groups comprise $50$\% CNOC2 group galaxies and $50$\%
   CNOC2 field galaxies ($\Pergr = 50$\%).} 
  \label{fig:ptrunclumfulldensevol}
\end{figure}

We adopt a realistic set of parameters governing the
spectrophotometric evolution, with a Kennicutt IMF, \zform = 10, solar
metallicity and a basic dust prescription \citep{Granato00}. The
progenitors of 2dF group galaxies are chosen by selecting all CNOC2
group galaxies plus a fraction of the CNOC2 field galaxies such that
the combined set is made up of $\Pergr$\% CNOC2 group members and
$(1-\Pergr)$\% CNOC2 field galaxies. As our field represents the
global population, the progenitors of 2dF field galaxies are simply
the CNOC2 field galaxies. Following the model prescription, we obtain
best fit values for $\ptrunc$ as a function of local luminosity and
environment to fit the observed evolution in $\fp$. Figures
~\ref{fig:ptrunclumnodensevol} and ~\ref{fig:ptrunclumfulldensevol}
show $\ptrunc$ as a function of luminosity in groups and the field
with $\Pergr=100$\% (local group members were all group members at
$z_{CNOC2}$) and $\Pergr=50$\% (local group members were $50$\% group
members and $50$\% field galaxies at $z_{CNOC2}$) respectively. We
note that adopting a Salpeter IMF does not significantly alter these
results.

\noindent Figures ~\ref{fig:ptrunclumnodensevol} and
~\ref{fig:ptrunclumfulldensevol} show the following:

\begin{enumerate}
  
\item{$\ptrunc$ is significantly greater than zero, implying that
    galaxy transformations are required over our redshift range, both
    in groups and the field. This is independent of the assumed
    evolution of clustering power and agrees with our conclusions in
    Section~\ref{sec-disquiesevol}. }
  
\item{Assuming no density evolution since $z \sim 0.45$, $\Pergr =
    100$\%, (Figure~ \ref{fig:ptrunclumnodensevol}), we see no
      evidence that $\ptrunc$ is larger in groups than in the field.
    This means that there must be some \emph{global} mechanism in
    which star formation can be effectively reduced to zero over a
    short period of time rather than simply declining in a quiescent
    manner as assumed in Section~\ref{sec-disquiesevol}. However the
    existence of a more evolved population (higher $\fp$) in groups
    suggests that the star formation history prior to $z \sim 0.45$
    must depend upon environment in some way. This could be either a
    nurturing environmental process at $z > 0.45$, or an
    earlier formation time for galaxies in groups (nature). We
    emphasize that our model is designed to simply match the observed
    evolution of $\fp$. It cannot simultaneously match evolution of
    the luminosity function, which requires a better understanding of
    the volume-averaged galaxy density. We are also constrained by our
    definition of ``field'' which spans the full range of environment.}
  
\item{If we assume a strong density evolution with only 50\% of local
    group galaxies in groups at $z \sim 0.45$, $\Pergr = 50$\%,
    (Figure ~\ref{fig:ptrunclumfulldensevol}), then a marginally
    larger $\ptrunc$ is invoked in groups than in the no density
    evolution ( $\Pergr = 100$\%) case, although not significantly so.
    Even at faint luminosities the differences between $\ptrunc$ in
    groups and the field is still of low significance ($< 2\sigma$ in
    the $-19.5 \leq$\MBj$\leq -18.5$ bin). 
    Physically, an
    enhanced $\ptrunc$ with greater density evolution is consistent
    with a second transformation process occuring during clustering as
    a galaxy is infalling into a larger dark matter halo. A strong
    density evolution with a $\ptrunc$ which remains constant with
    redshift could theoretically explain the larger value of $\fp$ in
    groups than in the field. However, realisations of dark matter
    halo merger trees suggest that the actual fraction of 2dF group
    galaxies in groups by $z = 0.45$ was $\sim 80$\% ($\Pergr = 80$\%)
    \citep{Lacey93}.}
  
\item{There are no clear trends of $\ptrunc$ with galaxy luminosity in
    groups. In the field there is a suggestion ($\sim 2\sigma$
    significance) that $\ptrunc$ decreases in the faintest bin ($-19.5
    \leq$\MBj$\leq -18.5$). 
}

\end{enumerate}

We acknowledge that our model is simple, and neglects the mass and
luminosity enhancing effects of galaxy-galaxy mergers. 2dFGRS studies of the local luminosity function, and its dependence upon environment and galaxy spectral type, suggest that galaxies with early spectral-types become more important in higher density regimes, particularly at low luminosities \citep{DePropris03,Croton04}. \citet{DePropris03} show that a simple model (similar to ours), in which star formation can be suppressed in clusters, can explain most of the differences between the cluster and field luminosity functions. They claim that mergers are only required to explain the small population of very bright early type cluster galaxies. 

The precise importance of galaxy mergers remains to be seen. There is significant evolution since $z \sim 1$ of the luminosity function of red galaxies in the COMBO-17 survey, and \citet{Bell04} conclude that mergers are required to explain at least some of this evolution. In clusters, comparisons of the K-band luminosity function over a similar redshift range suggest there is little evolution in the stellar mass of cluster galaxies \citep{dePropris99,KodamaBower03}. However, mergers are expected to be more common in groups than in clusters \citep[e.g.][]{Barnes85}. In a future paper, we will investigate the importance of mergers in groups, using existing data on CNOC2 and 2dF groups to map the evolution of the group K-band luminosity function. 

In a separate paper we will also make comparisons of the data with results from semi-analytic models of galaxy formation. Observational constraints on the bimodality of galaxy properties, and the dependence on galaxy luminosity, environment and redshift will place strong limitations on the physical processes regulating star formation in these models.

%We acknowledge that our model is simple, and neglects the mass and
%luminosity enhancing effects of galaxy-galaxy mergers. These are
%necessary to build up the population of bright galaxies in the local
%luminosity function which otherwise simply fade as they cease forming
%stars. In Paper~I we have shown that the CNOC2 groups have a relative
%excess of bright galaxies compared with the field, a trend which
%appears to hold in local groups (Figure~\ref{fig:lumfns}). This cannot
%be matched in the model without the incorporation of galaxy mergers.
%Incorporating mergers into our model can only serve to increase the
%requirement for transformations in the formation of locally bright
%galaxies from a fainter, more highly star forming population.
%We also note that in our
%model transforming galaxies tend to fade into fainter bins of
%luminosity, a process which affects the computation of $\ptrunc$. In a
%future paper we will make comparisons of the data with results from
%more physically realistic models\citep[e.g.][]{Cole00}.

\section{Conclusions}\label{sec-conclusions}

In this paper, we have examined the evolution of galaxies and the
effects of the group environment in kinematically selected groups from
the CNOC2 \citep[][supplemented with new and deeper Magellan
spectroscopy]{Carlberg01} and 2dFGRS \citep{Eke04} surveys. The data
span the redshift range $0.05 \leq z \leq 0.55$ and luminosities down
to \MBj $\leq -18.5$ (locally $\sim \Mstar+1.75$). Motivated by the
apparently fundamental differences between the blue, star-forming and
the red, passive populations of galaxies \citep{Balogh03,Blanton03} we
have arbitrarily divided our galaxies into \emph{passive}
(EW[OII]$<5$\AA) and \emph{star-forming} (EW[OII]$\geq 5$\AA)
populations. We have then shown that the fraction of passive
  galaxies $\fp$ is a strong function of:

\begin{itemize}
  
\item{{\bf redshift}: $\fp$ declines strongly with redshift, both in
    groups and the field and over the full luminosity range to at
    least $z \sim 0.45$. This is equivalent to a Butcher-Oemler
    trend in the emission line properties of group galaxies and in the
    global population.}
  
\item{{\bf environment}: $\fp$ is significantly higher in groups than
    the field across the full luminosity range, both locally and at $z
    \sim 0.45$.}
  
\item{{\bf luminosity}: $\fp$ increases steeply with luminosity across
    our range (\MBj$\lsim -18.5$) in groups and the field up to at
    least $z \sim 0.45$.}

\end{itemize}

Using the stellar population models of \citet{BC03}, we have shown
that the rate of evolution in $\fp$ since $z \sim 0.45$ cannot be
explained in a quiescent evolution scenario, i.e. by modelling
galaxies with a simple e-folding decline in their SFR. Even choosing
model parameters
geared to maximize this evolution cannot reproduce the observed difference
between 2dF and CNOC2 galaxies in $\fp$, especially fainter than
$\Mstar-0.5$.  This conclusion holds both in groups and the field.

We are therefore driven to assuming that transforming events take
place, in which star formation is abruptly
truncated, and have constrained the probability of truncation per
Gyr ($\ptrunc$) in groups and the field across the luminosity range
$-21.5 \leq $\MBj\ $\leq -18.5$. Although we have not
constrained the timescale of these events (simply assuming them to be
instantaneous), we show that their existence is strongly required by
the data ($\ptrunc \gg 0$). Surprisingly, we find no strong evidence
that $\ptrunc$ in the group environment exceeds that in the field. The
environmental dependence of $\fp$ requires that star formation history
prior to $z \sim 0.45$ must depend upon environment in some way. One
possibility is that as clustering of galaxies progresses an additional
suppression mechanism acts upon star forming galaxies as they fall
into groups (nurture). However, it is also possible to imagine
a nature scenario in which more strongly clustered galaxies
form first and all galaxies undergo transforming events, independently
of their environment. A better understanding of the environmental
influence on galaxy properties will be made possible by comparisons
with semi-analytic models, galaxies in other environments
\citep[e.g.][]{Nakata04} and higher redshift galaxy systems.

\section{Acknowledgements:}

We would like to thank the Magellan staff for their tremendous
support. RGB is supported by a PPARC Senior Research Fellowship. DJW,
MLB and RJW also thank PPARC for their support. VRE is a Royal Society
University Research Fellow. We are grateful to Dan Kelson for the use
of his spectral reduction software and to David Gilbank for his help
when learning to use it. We would like to thank Ian Lewis for his
measurements of EW[OII] in 2dFGRS. We also acknowledge Tom Shanks and
Phil Outram for their observations at Magellan and the full CNOC2,
2dFGRS and SDSS teams for outstanding datasets. Thanks also go to Bob
Nichol and Chris Miller for their help in producing the SDSS
catalogues. We thank Gustavo Bruzual and St\'ephane Charlot
for their publically available spectro-photometric evolutionary
modelling software GALAXEV and Carlton Baugh and Cedric Lacey for
helping to develop software used to model the galaxy properties we
required. Finally, we thank the anonymous referee for some 
useful feedback which has helped to improve this paper.

%\bibliography{mybib}

\appendix
\section{Aperture Effects:}\label{sec-apeffects}

The 2dF galaxies are observed spectroscopically through $2.1\arcsec$
diameter fibres at low redshift, corresponding to between $\sim
1.9$kpc ($z=0.05$) and $3.6$kpc ($z=0.1$). The CNOC2 galaxies are
observed through $1.3\arcsec$ (CNOC2) and $1.47\arcsec$ (LDSS2) slits
at much higher redshift corresponding to between $\sim 5.5$kpc and
$8.8$kpc. Therefore, significantly more flux will be lost from a large
galaxy in the low redshift 2dF sample than would be lost in its CNOC2
counterpart. The use of emission line equivalent widths rather than
fluxes reduces serious aperture effects by normalising to the
continuum level. However, a remaining worry is the existence of any bias towards
sampling primarily bulge light in the smaller apertures.
\citet{Baldry02} show that measurements of EW[OII] in 2dFGRS are
relatively insensitive to aperture size in repeat observations with
significant differences in seeing (differing by a factor of $>2$).
There is also no significant variation in the distribution of EW[OII]
over the redshift range $0.05 \leq z \leq 0.1$ over which we sample
little evolution but a factor of 2 in aperture diameter. However, the
variation in aperture size considered in this paper is somewhat larger
and so we have looked for clues in the SDSS for which resolved, digital
photometry exists.

Differences in selection method between SDSS and 2dFGRS are not
important when considering the aperture corrections to galaxies in the
same redshift range ($0.05 \leq z \leq 0.1$). \citet{Brinchmann03}
estimate aperture bias measurements of SFR per unit luminosity (SFR/L)
by constructing a likelihood distribution to determine the probability
of a given SFR/L for a given set of colours ((g-r),(r-i)) based on the
photometry within the fibre aperture. They then apply this likelihood
distribution to the galaxy population given the total galaxy
colours. The main assumption present in this technique is that the
distribution of SFR/L for a given colour is similar inside and outside
the fibre. However, we know that colour gradients can also be driven
by metallicity \citep{Hinkley01,Mehlert03,Tamura03}.
Therefore it is important that we understand the origin of the colour
gradients in the SDSS galaxy population before interpreting the level
of aperture bias in our data.

\begin{figure}

  \centerline{\psfig{figure=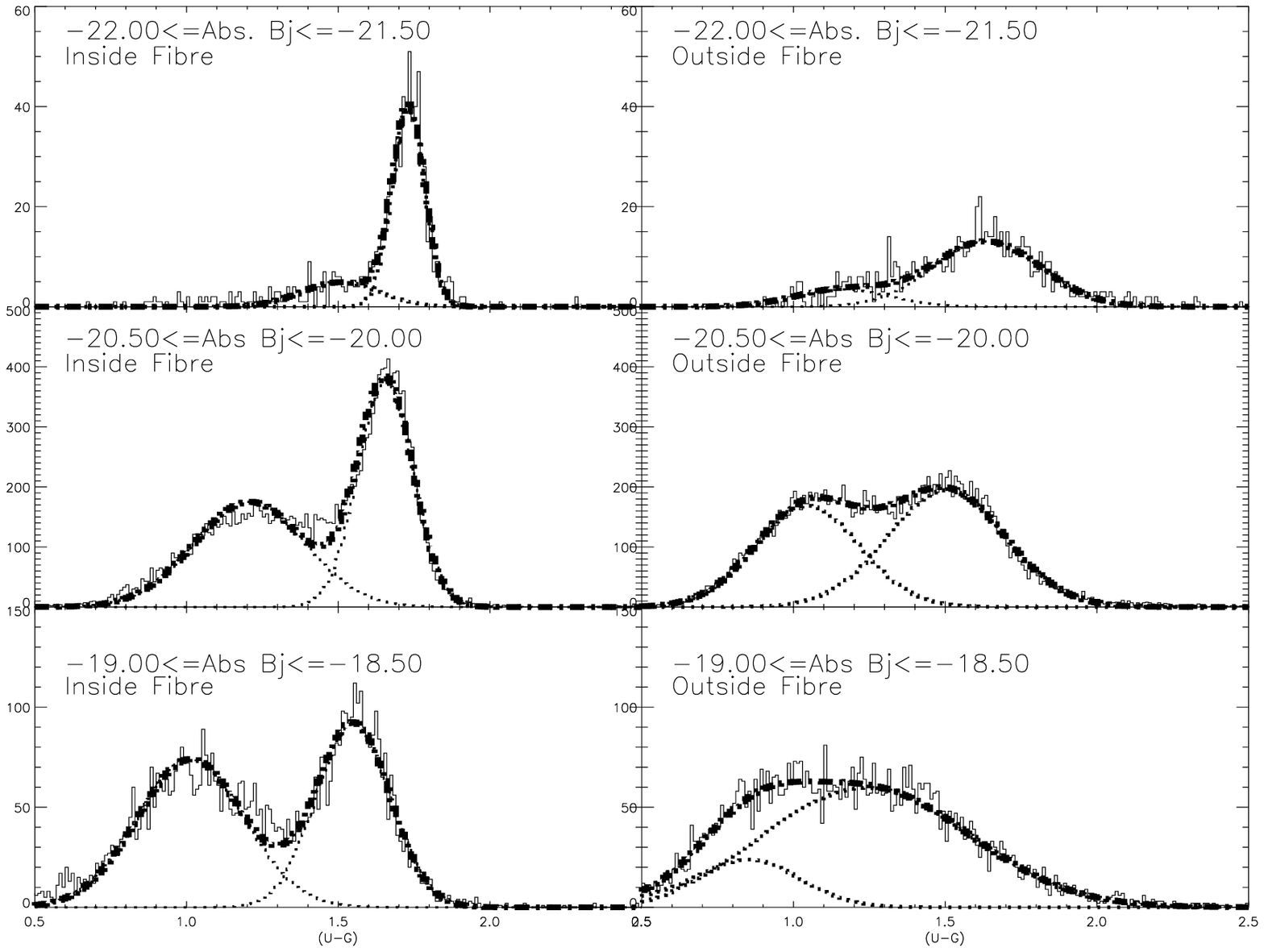,width=0.5\textwidth}}
 \caption{(u-g) colour distribution of SDSS galaxies {\it inside}
   (left) and {\it outside} (right) the fibre for {\it bright} ($-22
   \leq$\MBj$\leq -21.5$) galaxies (top), intermediate brightness ($-21
   \leq$\MBj$\leq -20.5$) and {\it faint} ($-19 \leq$\MBj$\leq -18.5$)
   galaxies (below). Each colour distribution is fitted by a double
   gaussian representing the bimodal populations of galaxies (thick
   line).} 
  \label{fig:SDSSapfitcols}
\end{figure}

We approximate the total galaxy colour of SDSS galaxies using their
Petrosian magnitudes. The fibre magnitudes measure the flux within a
SDSS spectroscopic fibre of diameter $3\arcsec$. We estimate the colour
of galaxies outside the fibre to be the Petrosian flux minus the
fibre flux. More details on the SDSS magnitude system can be found in
\citet{Stoughton02}. Whilst these fibres are larger than the 2dF
fibres, the poor seeing of 2dFGRS observations (a median seeing of
about $1.5\arcsec$) means that the 2dF fibres sample galaxy light from
a similar radius. We find that many SDSS galaxies with $0.05 \leq z
\leq 0.1$ do show significantly bluer colours outside of the fibre than
inside (see Figure~\ref{fig:SDSSapfitcols}).

We interpret the colours of galaxies in terms of the bimodal
distribution of red passive galaxies and blue star forming galaxies
\citep[as seen by][]{Balogh03,Blanton03,Baldry03}. \citet{Baldry03}
find that the colour distribution of the SDSS galaxy population is
well represented by a double gaussian model and so we choose a similar
method to fit the colour distribution of galaxies both inside and
outside the fibre. To make direct comparisons with our 2dFGRS galaxy
samples easier, we estimate the rest frame Petrosian $B_{j}$
band absolute magnitude of all SDSS galaxies in the range $0.05 \leq z
\leq 0.1$ using the transformation $b_{J} = g + 0.15 + 0.13(g-r)$
where g and r have been k-corrected \citep{Norberg02}. The double
gaussian model is fit to the galaxy population using a
gradient-expansion algorithm to compute a non-linear least squares
fit. Fits to 3 of these bins of luminosity, spanning the full
significant luminosity range can be seen in
Figure~\ref{fig:SDSSapfitcols}. The wider peaks seen outside the fibre
can be attributed to the measurement errors on the galaxy magnitudes.
These errors are roughly twice as large in computed magnitudes outside
the fibre than in fibre magnitudes. Fainter than \MBj$ = -20$, the
median measurement error of $\gsim 0.15$mag outside the fibre (and up
to $\sim 0.5$mag in some galaxies) smoothes out the double gaussian
distribution, as can be seen in the bottom-right panel of
Figure~\ref{fig:SDSSapfitcols}. The double gaussian fit to the colour
distribution is then poorly constrained. Therefore, we only consider
the galaxy population brighter than \MBj$ = -20$ in this analysis
(aperture effects should be less important for the less luminous
galaxies, anyway).

\begin{figure}

  \centerline{\psfig{figure=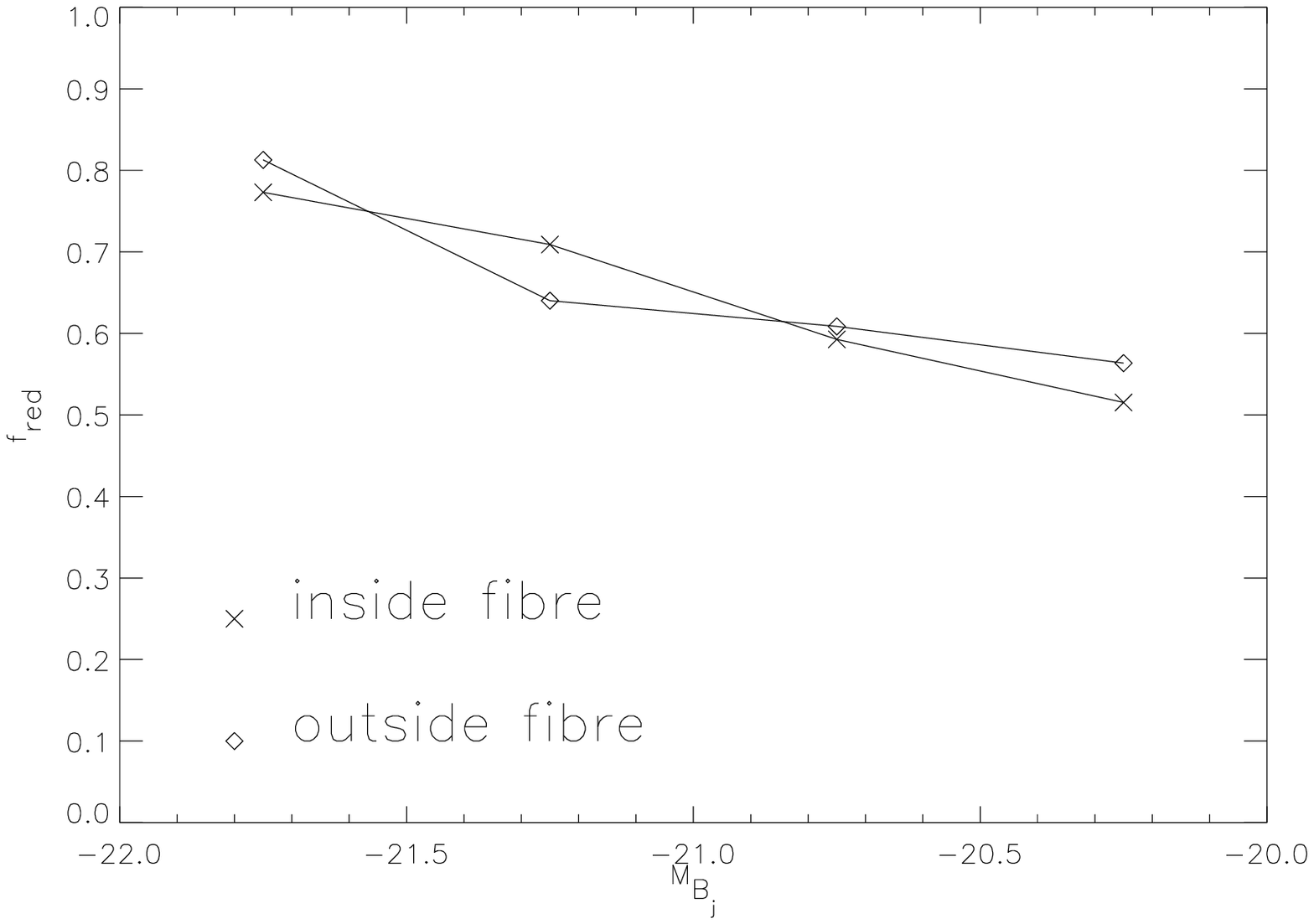,width=0.45\textwidth}}
\caption{The fraction of galaxies located in the red peak of the
  bimodal distribution ($\fred$) as a function of luminosity, both
  inside and outside the fibre. The crosses represents the fibre
  colour distribution and the diamonds the colour distribution
  outside the fibre.
} 
\label{fig:SDSSfracredmagfibout}
\end{figure}

\begin{figure}

  \centerline{\psfig{figure=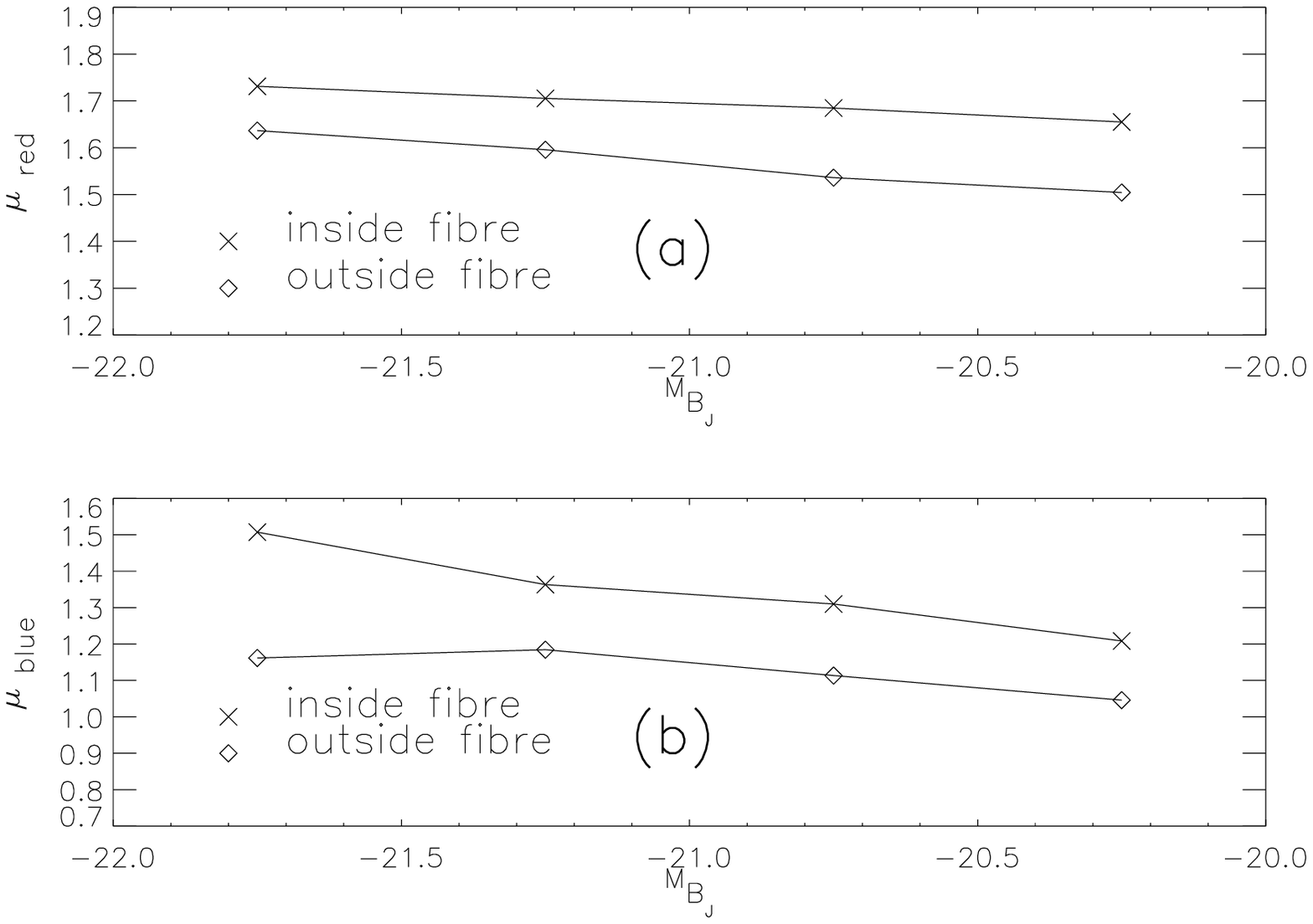,width=0.45\textwidth}}
\caption{The central position (mean) of the red peak ({\bf a}, $\mu_{red}$) and the blue peak
  ({\bf b}, $\mu_{blue}$) of the bimodal distribution as a function of luminosity, both
  inside and outside the fibre. The crosses represents the fibre
  colour distribution and the diamonds the colour distribution
  outside the fibre.
} 
\label{fig:SDSSmeanredmagfibout} 
\end{figure}

Figures~\ref{fig:SDSSfracredmagfibout} and
~\ref{fig:SDSSmeanredmagfibout} respectively, show the variation with
luminosity, inside and outside the fibre, of the fraction of galaxies located inside
the red peak ($\fred$), and the mean (u-g) colour of
the red peak ($\mu_{red}$) and blue peak ($\mu_{blue}$).
In particular, Figure~\ref{fig:SDSSfracredmagfibout} shows that the fraction of
galaxies located in the red peak $\fred$ outside the fibre is
consistent with $\fred$ inside the fibre. This indicates that no
  aperture corrections are necessary to account for the fraction of
  red, passive galaxies in the sample.

Figure~\ref{fig:SDSSmeanredmagfibout} shows a
bluewards shift of $\mu_{red}$ as we move to fainter magnitudes both
inside and outside the fibre. However, a radial colour gradient exists
at all luminosities, such that $\mu_{red}$ is $\sim 0.1-0.15$
magnitudes bluer in the outer regions. A comparable colour gradient is
seen in the blue peak (the blue population of galaxies). 
The similarity of the colour gradient in both the blue galaxies and in the
red, passive galaxies (in which no star formation is expected) suggests 
that it may arise from a metallicity gradient rather than an age
gradient, and explains
why we observe no trend in $\fred$ with aperture.
This interpretation is supported by the observations of metallicity
gradients (and the lack of age gradients) in early-type galaxies
\citep[e.g][]{Hinkley01,Mehlert03,Tamura03,Wu04}.
The colour differences we observe inside and outside the fibre are $\delta(u-g) \sim
0.15$, consistent with the average (u-g) colour gradient of 0.18
found in 36 early type SDSS galaxies analysed by \citet{Wu04}. 

\section{Simple Models of Galaxy Evolution}\label{sec-appmodels}

\subsection{The Quiescent Evolution Scenario}\label{sec-quiesevol}

A quiescent evolution scenario is characterized by the lack of sudden
events which drastically alter a galaxy's star formation. In this
scenario, the star formation rate (SFR) in any galaxy declines
with an e-folding timescale, $\tau$. This timescale is
short in the case of massive early-type galaxies, and much longer in
the case of later types. The environmental dependence of star
formation can then be invoked using a nature-origin scenario in
which more early-type galaxies form in more densely clustered regions
of the Universe.

To test whether this model can explain the strong evolution seen in
our data, we must first model the ways in which galaxy luminosity and
EW[OII] depend upon the star formation history of a galaxy. We do this
by modelling the spectrophotometric evolution of CNOC2 galaxies
\citep{BC03} with different forms of star formation history. By
accounting for this evolution, we can understand how $\fp$ in CNOC2
galaxies can be compared with the equivalent values of $\fp$ locally
in the 2dF data. This model also requires no density evolution which
means that group galaxies remain as group galaxies and field galaxies
remain as field galaxies. The model evolution contains the parameters
[\emph{IMF,\zform,Z,dust?}] and is applied in the following way:

\begin{enumerate}
  
\item{Each model galaxy is given an IMF, redshift of formation,
    \zform, characteristic timescale, $\tau$ and metallicity Z. We
    also choose either a model with no dust or with a
    \citet{Granato00} Milky Way dust extinction law applied.}
  
\item{\citet{BC03} model SEDs are used to model the spectrum of a
    galaxy with the chosen parameter set at various steps in redshift
    up to $z=0.55$.}
  
\item{The rest-frame \Bj-band luminosity evolution between two
    different redshifts is modelled by convolving the filter
    transmission function with the model spectrum (normalised to a fixed
    stellar mass) at each redshift and calibrating to Vega as in
    2dFGRS.}
  
\item{The evolution of EW[OII] is measured by computing the model
    Lyman continuum flux in each spectrum and artificially
    reprocessing this as [OII] flux using the HII region models of
    \citet{Stasinska90} at the chosen metallicity, Z. We assume 1
    ionising star per HII region with effective temperature 45000K and
    a HII region electron density of $10$cm$^{-3}$. The equivalent
    width is then simply measured by computing the continuum
    luminosity at the wavelength of the [OII] emission line and
    normalising the line flux by its continuum level. We have
    successfully tested our model by reproducing the results of
    \citet{PoggBarb96} for an elliptical galaxy with a recent
    starburst.}

\end{enumerate}

At a given redshift and for a given IMF, Z, \zform\ and dust option,
we can determine a value of $\tau$ at which EW[OII] = 5\AA. By
measuring $\tau = \tau_{lim}$ at low redshift (in the 2dF redshift
range), we can then determine the equivalent value of EW[OII] for the
same galaxy (with $\tau = \tau_{lim}$) at higher redshift (i.e. at
CNOC2 redshifts). This value we then call $x$, in units of \AA.

In this way, we determine the dependence of $x$ on all the
relevant parameters. Higher values of $x$ imply greater evolution in a
galaxy's SFR and so the most extreme example of quiescent evolution
will occur with a parameter set in which $x$ is chosen to be as large
as realistically possible:

\begin{itemize}
  
\item{$x$ is approximately $1/3$ larger for a Salpeter IMF than for a
    Kennicutt IMF \citep{Kennicutt83}.}
  
\item{$x$ decreases when dust is included.}
  
\item{$x$ is at a peak where the metallicity, Z is approximately solar
    ($=0.02$) or slightly sub-solar (down to $Z \sim 0.004$). Both at
    lower and higher metallicities, the value of $x$ decreases.}
  
\item{On the whole, $x$ decreases as \zform\ increases.}
  
\item{$x$ generally increases for larger choices of CNOC2 redshift
    (i.e. up to $z = 0.55$).}
  
\item{$x$ increases for lower choices of 2dFGRS redshift (i.e. down to
    $z=0.05$).}

\end{itemize}

The rest-frame \Bj-band luminosity evolution of a CNOC2 galaxy is
computed by determining the value of $\tau$ which best reproduces the
value of EW[OII] for that galaxy at $z_{CNOC2}$. Then the fading of
that galaxy by $z_{2dF}$ is computed using $\delta$\MBj =
\MBj$(z_{2dF})$ - \MBj$(z_{CNOC2})$ for those model parameters.

The mean redshifts of the 2dFGRS and CNOC2 samples are $z \sim 0.08$
and $z \sim 0.45$ respectively. The values of $\tau_{lim}$ and $x$ in
our 2 test cases (see main text) are then:
\begin{itemize}
\item{{\bf Control model:} With a Kennicutt IMF, \zform = 10 and solar
    metallicity and the \citet{Granato00} dust prescription,
    $\tau_{lim}=2.93$Gyrs and $x=7.04$\AA.}
\item{{\bf Extreme model:} With a Salpeter IMF, \zform = 3, solar
    metallicity and no dust, $\tau_{lim}=1.74$Gyrs and $x=9.86$\AA.}
\end{itemize} 

\subsection{The Truncation Scenario}\label{sec-truncevol}

In the truncation scenario, we allow galaxy transformations to occur
in which a galaxy's star formation drops instantaneously to zero. This
acts as a simple way to enhance the decline of star formation, and in
particular to turn a star forming galaxy into a passive galaxy,
independent of its initial star formation rate. In reality such
transformations may be accompanied by a strong starburst phase or/and
a longer timescale decline to zero star formation. However a
detailed modelling of spectral and photometric parameter space would
be necessary to constrain these elements of the model with enough
accuracy. In this paper, we concentrate on matching the value of $\fp$ as
defined using the value of EW[OII] over a range of \MBj\ luminosity.
This is enough information to constrain the probability of
transformations using a simple model, similar to our quiescent
evolution model described in Section~\ref{sec-quiesevol}.

The probability of truncation ($\ptrunc$) is constrained as a function
of local luminosity in groups and the field by randomly choosing an
evolution to $z_{2dF}$ of star formation for each CNOC2 galaxy with a range of
truncation probabilities. The CNOC2 galaxies are then evolved
appropriately in \MBj\ and EW[OII] using \citet{BC03} models and the
resulting $\fp$ for the evolved population is compared with the local
values obtained from 2dF data, thus constraining $\ptrunc$. Density
evolution is incorporated by requiring local groups to contain
$\Pergr$\% CNOC2 group members and $(1-\Pergr)$\% CNOC2 field
galaxies. The model contains the spectrophotometric evolution
parameters [\emph{$\ptrunc$,IMF,\zform,Z,dust}] and the density
evolution parameter $\Pergr$. However, this is simplified by
maintaining a consistent and reasonable spectrophotometric model. We
choose a Kennicutt IMF, redshift of formation, \zform\ = 10.0 and
solar metallicity. We also incorporate a constant dust prescription in
the model \citep{Granato00}. We note from experimentation that
changing these parameters does not strongly affect our conclusions
(dependencies on these parameters can be seen in the
Section~\ref{sec-quiesevol}). Our model is implemented as follows:

\begin{enumerate}
  
\item{A fiducial set of model parameters is chosen. These include IMF,
    redshift of formation, \zform\, metallicity Z and presence (or
    not) of dust extinction.}
  
\item{For the chosen set of parameters, galaxy spectra are constructed
    for a variety of star formation histories, using \citet{BC03}
    model SEDs. The star formation histories are parameterized with
    characteristic timescale $\tau$ and redshift of truncation,
    $\ztrunc$ allocated via a 2D grid of discrete values for ease of
    computation. We compute histories combining $\tau$ = [1000, 15,
    12, 10, 9, 8, 7, 6, 5, 4, 3, 2, 1, 0.5] Gyrs and $\ztrunc$
    corresponding to 12 equally spaced intervals in time between $z =
    0.55$ and $z = 0.05$, given our cosmology. One set of models
    histories with no truncation is also computed.}
  
\item{The evolution in rest-frame \Bj-band luminosity and EW[OII] are
    computed using the same method as described in the quiescent
    evolution model.}
  
\item{At the redshifts z = [0.55, 0.5, 0.45, 0.4, 0.35, 0.3, 0.08] we
    determine the values of rest-frame \Bj-band luminosity per unit
    stellar mass, EW[OII] and the ratio of stellar mass to present day
    stellar mass for all possible combinations of $\tau$ and
    $\ztrunc$. This covers the CNOC2 redshift range and the mean 2dF
    redshift ($z_{2dF}=0.08$).}
  
\item{For each star formation history (i.e. each value of $\tau$ and
    $\ztrunc$), we compute the evolution in \MBj\ and EW[OII] from
    $z_{CNOC2}$ = [0.55, 0.5, 0.45, 0.4, 0.35, 0.3] to $z_{2dF}=0.08$.
    For intermediate $z_{CNOC2}$ we simply interpolate between these
    values.}
  
\item{For each galaxy in the CNOC2 sample, a value of $\tau$ is chosen
    which best matches the EW[OII] of the CNOC2 galaxy at the redshift
    of that galaxy. This involves making a 2D interpolation over the
    $\tau$ models in EW[OII] and $z_{CNOC2}$.}
  
\item{The probability of a galaxy having its star formation truncated
    in 1 Gyr is $\ptrunc$. For a given value of $\ptrunc$, the CNOC2
    galaxies are evolved to redshift $z = 0.08$ for comparison with
    2dF galaxies. This evolution consists of randomly selecting a
    truncation redshift, $\ztrunc$, where $z_{CNOC2} > \ztrunc > 0.08$
    with a probability equivalent to the product of $\ptrunc$ and the
    timestep in Gyr,
    for each $\ztrunc$. Each galaxy can only experience one truncation
    and if it has not undergone any truncation by $z=0.08$ then we
    select the evolution model with no truncation. The values of \MBj\ 
    and EW[OII] for the evolved CNOC2 galaxy at $z=0.08$ are then
    assigned in a consistent manner from the computed evolution
    models.}
  
\item{This evolution is repeated with a range of values of $\ptrunc$
    to create a series of mock catalogues of CNOC2 galaxies evolved to
    $z=0.08$. We allow $\ptrunc$ to vary between 0.0 and 0.8 in steps
    of 0.005.}
  
\item{A density evolution model is assumed. In this model, CNOC2 field
    galaxies become 2dF field galaxies and CNOC2 group galaxies become
    2dF group galaxies. However, a CNOC2 field galaxy may also become
    a 2dF group galaxy, with a probability $\pfldgrp$ (which is
    computed such that local groups comprise $\Pergr$\% CNOC2 group
    members and $(1-\Pergr)$\% CNOC2 field galaxies), mimicking the
    clustering of large scale structure in the Universe. Realisations
    of dark matter halo merger trees suggest that the actual fraction
    of 2dF group galaxies in groups by $z = 0.45$ was $\sim 80$\%
    ($\Pergr = 80$\%) \citep{Lacey93}}
  
\item{Given our choice of density evolution, for each evolved CNOC2
    mock catalogue (each choice of $\ptrunc$) we compute $\fp$(mock)
    as a function of luminosity in the group and field samples. These
    values are then compared with the locally measured values
    $\fp$(2dF) and a best fit value of $\ptrunc$ is chosen for each
    luminosity bin in each sample using a polynomial function to fit
    $\fp$(evolved CNOC2) as a function of $\ptrunc$.}
  
\item{Errors on $\ptrunc$ are determined by first determining the
    errors in $\fp$(mock) and $\fp$(2dF) and then combining these in
    quadrature and converting to an error in $\ptrunc$ in each bin.
    Errors in $\fp$(mock) include the statistical errors in the CNOC2
    population and its evolution and the error in EW[OII] leading to
    an error in the distribution of $\tau$ models selected. Errors in
    $\fp$(2dF) include the statistical error in the population and the
    error due to random smoothing (by 2\AA) of the 2dF galaxies'
    EW[OII]. These errors are all estimated using a resampling
    method.}

\end{enumerate} 

\end{document}